\documentclass[preprint,3p,10pt]{elsarticle}

\usepackage{graphicx}
\usepackage{amssymb}
\usepackage{lineno}
\usepackage{booktabs}

\usepackage{float}

\usepackage{hyperref}
\hypersetup{
  colorlinks=true,
}

\journal{arxiv}

\begin{document}
\hypersetup{
  linkcolor=red,
  urlcolor=blue,
  citecolor=red
}

\begin{frontmatter}

%% Title, authors and addresses

\title{An Empirical Analysis of the R Package Ecosystem}

%% use the tnoteref command within \title for footnotes;
%% use the tnotetext command for the associated footnote;
%% use the fnref command within \author or \address for footnotes;
%% use the fntext command for the associated footnote;
%% use the corref command within \author for corresponding author footnotes;
%% use the cortext command for the associated footnote;
%% use the ead command for the email address,
%% and the form \ead[url] for the home page:
%%
%% \title{Title\tnoteref{label1}}
%% \tnotetext[label1]{}
%% \author{Name\corref{cor1}\fnref{label2}}
%% \ead{email address}
%% \ead[url]{home page}
%% \fntext[label2]{}
%% \cortext[cor1]{}
%% \address{Address\fnref{label3}}
%% \fntext[label3]{}

%% use optional labels to link authors explicitly to addresses:
%% \author[label1,label2]{<author name>}
%% \address[label1]{<address>}
%% \address[label2]{<address>}

\author{Ethan Bommarito}
\ead{ethan@licens.io}
\author{Michael J Bommarito II}
\address{Licensio, LLC}
\ead{mike@licens.io}

\begin{abstract}
In this research, we present a comprehensive, longitudinal empirical summary of the R package ecosystem, including not just CRAN, but also Bioconductor and GitHub. We analyze more than 25,000 packages, 150,000 releases, and  15 million files across two decades, providing comprehensive counts and trends for common metrics across packages, releases, authors, licenses, and other important metadata.  We find that the historical growth of the ecosystem has been robust under all measures, with a compound annual growth rate of 29\% for active packages, 28\% for new releases, and 26\% for active maintainers. As with many similar social systems, we find a number of highly right-skewed distributions with practical implications, including the distribution of releases per package, packages and releases per author or maintainer, package and maintainer dependency in-degree, and size per package and release.  For example, the top five packages are imported by nearly 25\% of all packages, and the top ten maintainers support packages that are imported by over half of all packages.  We also highlight the dynamic nature of the ecosystem, recording both dramatic acceleration and notable deceleration in the growth of R.  From a licensing perspective, we find a notable majority of packages are distributed under copyleft licensing or omit licensing information entirely.  The data, methods, and calculations herein provide an anchor for public discourse and industry decisions related to R and CRAN, serving as a foundation for future research on the R software ecosystem and ``data science'' more broadly.
\end{abstract}


\end{frontmatter}

%% main text
\section{Introduction}
\label{S:introduction} 

Since its first release in 1993, R has established itself as the most popular open-source statistical computing language  \cite{tiobe2020r}.  One common explanation for this success is R's large community, partially inherited from S, the language it ``succeeded,'' and its rich ecosystem of open-source packages and contributors.  While its relative popularity has varied over the years, the increasing emphasis on statistical analyses in academia and industry has been evidenced by R's increasing absolute rank in the TIOBE index.  Inspired by research such as \cite{decan2015development} \cite{decan2016github} \cite{decan2016topology} \cite{decan2017empirical} \cite{decan2019empirical}, our prior research \cite{bommarito2019empirical} on Python's PyPI, and professional experience with software development and information security, we seek in this paper to empirically describe the package ecosystem of this important language.  Unlike extant literature, we analyze both complete historical package metadata and package contents, providing a more comprehensive understanding of releases, authors, licenses, dependencies, and other trends in package source and metadata over time. This research is intended to provide a convenient reference for empirical claims regarding the R ecosystem and to provide an anchor for a larger body of future research.

R's most well-known source of packages, the Comprehensive R Archive Network (CRAN), was first proposed in 1996 by Kurt Hornik and collaborators at TU Wien after inspiration by CTAN and CPAN.  They announced the first realization of this proposal in early 1997 and launched the original server in the same year \cite{hornik2002vienna}.  Since then, the number of ``official'' CRAN mirrors has grown to over 100 in 2020 \cite{cranstatus}.  As its official and longest-serving package repository, CRAN provides an empirical source of information about both R specifically and ``data science'' broadly.  While a number of other studies have examined CRAN, either alone or in combination with GitHub, these studies have generally relied on smaller samples, metadata-only information, and qualitative coding in their analysis \cite{MORACANTALLOPS2020110744} \cite{mora2020evolution} \cite{decan2016github} .  While our research does not address the same questions as these publications, it does establish a complete, longitudinal, and reproducible baseline for CRAN across its history, including direct analysis of package source code and data.

Over the two decades since CRAN's launch, a number of academic communities have especially embraced R.  Most notably, many members of the bioinformatics field have standardized on the usage of R in their work.  In 2001, a group of such researchers began the ``Bioconductor project[...], an initiative for the collaborative creation of extensible software for computational biology and bioinformatics (CBB)''.  Their own words exemplify why so many researchers have been drawn to R: ``[t]he primary motivations for an open-source computing environment for statistical genomics are transparency, pursuit of reproducibility and efficiency of development'' \cite{gentleman2004bioconductor} \cite{bioconductor2020}.  Like CRAN, the Bioconductor project provides its own repository of packages managed and distributed based on the needs of the bioinformatics community.  Similar projects, such as Omegahat, have also intermittently existed over the last two decades \cite{omegahat2020}.  Despite its prominent role in an important area of research, Bioconductor has received far less attention in prior research outside of \cite{nagarajan2013packdep} and \cite{decan2015development}, both of which are dated at this point.

In the context of society broadly, the last five years have witnessed a dramatic increase in attention towards the ``emerging'' fields of data science, machine learning, and ``AI.''  R has long held a prominent position in the endeavors of academic and industrial researchers in this space.  As statistics departments often transitioned their primary teaching materials from S to R, many of the university students who filled the early ranks of ``data scientists'' were trained in R as their primary language.  In industry, organizations including Microsoft, Google, Oracle, Facebook, and IBM have been active in the area, embedding R in databases and BI platforms, acquiring companies that provide commercial support and extension, publishing their own R packages, and collaborating with core developers.  Many of these academic and industry activities have also been open source, and, as is common among open-source activities, have been recorded on GitHub \cite{github2020}.  GitHub itself has become increasingly popular over this time, and the introduction of the \textsc{devtools} package for R has greatly simplified the process of developing and using packages from GitHub. 

Together, CRAN, Bioconductor, and GitHub contain a record of the evolving activities of authors and maintainers as they adapt to and affect the environment they co-create - metaphorically, an ecosystem.  Each package source offers different benefits to package developers and package users.  As a purpose-built distribution, Bioconductor offers members of the bioinformatics community a focused and well-tested platform for research, including both original research and subsequent replication.  This platform is the result of a committee of experts and developers who are constantly supporting and improving the distribution, though at their sole discretion and on their own release schedule.  In contrast, while CRAN provides official packages and some degree of quality control, the CRAN repository is more open than Bioconductor.  Packages of any nature may be submitted to CRAN so long as they comply with its policies and pass its basic automated test suite.  These packages are made available as soon as approved by the responsible team of R maintainers or community members.  GitHub lies at the most open end of the spectrum, containing not just packages meant for re-use, but also any source code, data, or related activities that an individual or organization makes available. Many package developers who release in CRAN or Bioconductor use GitHub to manage their development activities prior to ``official'' package or distribution release. The \textsc{devtools} package even allows package authors to distribute their packages directly from GitHub without relying on either the Bioconductor or CRAN teams for review or approval.

\section{Data and Methods}
\label{S:data}

Decades have passed since the R language and CRAN first appeared, and many things have changed since their first release and deployment.  The data presented in this research is based on a platform developed by the authors to archive and analyze common software languages for the purpose of information security, compliance, code quality, and valuation \cite{licensio2020}.  With respect to R, the platform's data retrieval protocols are outlined below:

\subsection{CRAN Data Retrieval}
\label{S:cran_data_retrieval_procedure}
\begin{enumerate}
    \item Retrieve a list of all CRAN packages from a trusted mirror \small{(e.g., \url{https://cloud.r-project.org/})}\footnote{Most mirrors include sub-folders for ``archived,'' ``abandoned,'' or ``old(er)'' packages.  We retrieve these packages, even though they may not be currently available to install.}
    \item For each package $P$,
    \begin{enumerate}
        \item Download most recent release from \small{\url{/src/contrib/}}
        \item Download all archived releases of $P$ from \small{\url{/src/contrib/Archive/P}}
        \item For each release $R$ in package $P$,
        \begin{enumerate}
            \item Parse and store the release metadata in the DESCRIPTION file
            \item Parse and store any license or copyright information in the LICENSE file
            \item For each file $F$ in release $R$,
            \begin{enumerate}
                \item Analyze and index $F$ in file database
                \item If $F$ is classified as a source file, calculate source code metrics
            \end{enumerate}
        \end{enumerate}
    \end{enumerate}
\end{enumerate}

In many cases, source and data files do not change from release to release.  In the extreme case, releases may only update package structure or metadata.  We perform SHA-1 hashing of all files to efficiently index, reduce unnecessary reprocessing, and identify identical files across all packages and releases.

\subsection{Bioconductor Data Retrieval}
\label{S:bioc_data_retrieval_procedure}
\begin{enumerate}
    \item Retrieve a list of Bioconductor versions from the Bioconductor server
    \item For each Bioconductor release version $V$ greater than 2.5,\footnote{Due to issues with historical Bioconductor package listings, retrieval of historical Bioconductor packages prior to version 2.5 produced inconsistent and incomplete results.  As a result, they have been excluded from this analysis.}
    \begin{enumerate}
        \item Retrieve a list of all package releases from \small{\url{https://bioconductor.org/packages/V/bioc/}}
        \item For each release $R$,
        \begin{enumerate}
            \item Parse and store the release metadata in the DESCRIPTION file
            \item Parse and store any license or copyright information in the LICENSE file
            \item For each file $F$ in release $R$,
            \begin{enumerate}
                \item Analyze and index $F$ in file database
                \item If $F$ is classified as a source file, calculate source code metrics
            \end{enumerate}
        \end{enumerate}
    \end{enumerate}
\end{enumerate}

\subsection{GitHub Data Retrieval}
\label{S:github_data_retrieval_procedure}
\begin{enumerate}
    \item Using the GitHub Search API, collect all releases that are tagged under the R language\footnote{The GitHub Search API does not return stable results, requires rate limiting, and has peculiar chunking requirements.  Our search results were performed by adaptively chunking requests by push date and size, but results may vary based on chunking strategy.}
    \item For each release $R$,
    \begin{enumerate}
        \item Parse and store the metadata returned from the GitHub Search API
        \item Parse and store the release metadata in the DESCRIPTION file
        \item Parse and store any license or copyright information in the LICENSE file
        \item For each file $F$ in release $R$,
        \begin{enumerate}
            \item Analyze and index $F$ in file database
            \item If $F$ is classified as a source file, calculate source code metrics
        \end{enumerate}
    \end{enumerate}
\end{enumerate}

As the protocol descriptions imply, data is collected from CRAN and Bioconductor in a similar manner. Current releases are available under a distribution folder on a mirror, and historical releases are grouped by package folder containing one or more prior releases.  Whereas CRAN is organized by R version, Bioconductor is organized by its own package releases, live or archived, tied to a Bioconductor version.  

GitHub, however, differs from this structure quite radically, and the nature of GitHub repositories makes the data much less simple.  In this analysis, we analyze releases as published in GitHub repositories, which represent the most comparable record to CRAN and Bioconductor. Notably, GitHub repositories also include GitHub metadata, e.g., license type, not just \textsc{DESCRIPTION} metadata.  This GitHub release functionality was made generally available by GitHub in 2013, meaning that packages that were distributed through GitHub prior to 2013 are not captured in our results below.  As \textsc{devtools} 1.0 was not released until 2013, we believe that only a small number of packages or releases are omitted.  While \textsc{devtools} supports installing from commits, tags, branches, pull requests, or releases, the analysis of commit-by-commit, branch-by-branch development practices is not our focus in this research. In the instance where the metadata GitHub provides is in conflict with the metadata provided in a DESCRIPTION file, we prefer the information provided in the DESCRIPTION file, since it is more likely to agree with other versions of that package hosted on other sites.

Note that the retrieval methods above retrieve releases from three separate sources of R packages. While many packages or releases are only available from one archive, some packages or releases may be present in more than one source.  For example, many authors may release a tagged version on GitHub and submit the GitHub archives to CRAN for listing.  That package may subsequently be distributed as part of a Bioconductor release.  In some cases, identical sources may be present with varying versions or metadata, e.g., to explicitly indicate compatibility with a newer R version.  So as to avoid overcounting of such identical or trivially-varying releases, all counts presented in this research reflect deduplication of releases for the same package name and version string or SHA-1 hash.  For example, while \textsc{ggplot2 2.2.1} exists as both a GitHub release and on CRAN, we count it as a single unique package release for 2016.\footnote{Outside of counts, these releases are analyzed separately, e.g., for information security issues, so long as they are not identical by SHA-1 hash.}  Further, much of the activity on GitHub may relate to forks of popular packages as users contribute via GitHub pull request workflows.  Such activities are an important part of a healthy open-source community, but may result in overcounting  R activity at the package level.  On the other hand, we also rely on GitHub's R tagging and release functionality for this search; authors who never use the GitHub release functionality or whose source is not properly tagged by GitHub are absent from our results, and this may result in undercounting R activity at the package and author level.  In our analysis of forks with releases, we find only 42 packages occur with this condition, and so we believe the results do not materially impact our interpretations.  Further, many users fork R packages without ever committing a modification.  By only including GitHub releases, we exclude these trivial forks, which would otherwise outnumber real packages. 

Lastly, a number of important metadata fields demonstrate great variance across contributors and within contributors.  Contributors record their name with and without initials, middle name, or credentialing.  Contributors include, omit, or change their email addresses or affiliations.  Contributors use different Relator values over time.  Similar dynamics occur across many other fields.  We perform simple normalization for authors, relying on both name and email address, and have manually reviewed these normalizations.  However, normalization of such inconsistencies in other fields is left to future work on specific areas of interest.

\section{Results}
\label{S:results}
As described above in Section \ref{S:data}, our results are based on three sources of R packages in the ecosystem: CRAN, Bioconductor and GitHub. CRAN is the most well-known and longitudinal among these, as it is the largest, oldest, and best-known source. GitHub, while not exclusive to R or any other language, is immense compared to CRAN in the terms of the number of authors, the amount of activity, and how much R source code is published there.  Relative to these two sources, Bioconductor is the most focused and smallest. Despite these large variations in scale between the sources, we attempt to present them in parallel where possible in the tables and figures below.  The majority of our tables below are grouped by source, year, or another similar index. Where possible, we standardize the presentation of results on the ``package'' or ``release'' level (a ``package'' is a grouping of one or more ``releases'').  We track package releases across sources, so if the same package version is released across more than one source, we can keep their releases separated by source in the event that identical versions do not contain identical contents.  

While the authors' platform executes these procedures on a daily basis for commercial purposes, the tables and figures presented herein are based on a snapshot from December 2020.  A summary of data is provided in Table \ref{tab:ecosystem_summary}, indicating the number of records in our platform as of this time.  In total, we analyze over approximately 15 million files across over 150,000 releases from over 15,000 authors.  The unique license metric counts the number of unique metadata values, and its magnitude reflects the fact that there is no validation for the license fields in \textsc{DESCRIPTION} files; some authors use CRAN-specific or SPDX standards, but many do not.  We discuss future work on this topic in Section \ref{S:conclusion} below.

\begin{table}[htbp]
    \centering
    \begin{tabular}{r|r|r|r}
        \toprule
        \textbf{Statistic} &    \textbf{CRAN} &   \textbf{Bioconductor} &   \textbf{GitHub}\\
        \midrule
        Number of Packages    &  20,023 &  2,043 &  6,650 \\
        Number of Releases    & 115,134 & 21,335 & 19,225 \\
        Number of Maintainers &  14,131 &  1,957 &  1,432 \\
        Number of Unique Licenses   &  609 &  109 &  284 \\
        Number of Files & 11,713,946 & 1,815,723 & 4,124,235  \\
        \bottomrule
    \end{tabular}
    \caption{Summary of key statistics by source}
    \label{tab:ecosystem_summary}
\end{table}

We divide the remainder of the results section as follows.  First, in Subsection \ref{s:results_growth}, we present time series of counts for packages, releases, and maintainers over time, allowing us to summarize the growth of R over the last two decades.  Next, in Subsection \ref{s:results_packages}, we examine key facts and distributions related to packages and releases \textit{per se}, such as the number of packages over time and the distribution of duration between release.   In Subsection \ref{s:results_contributors}, we present information about contributors such as authors and maintainers, including measures of the most active maintainers.

\subsection{Growth}
\label{s:results_growth}
While these total counts are impressive, it is important to understand how the ecosystem together and sources alone have grown over time to create today's collection of software.  Importantly, our analysis window for each of these sources varies.  While we have archived all CRAN releases for two decades, our reliable data for Bioconductor and GitHub is limited to only the past decade.  As such, we will initially present and interpret their time series independently.

Table \ref{tab:cran_time_series} documents the number of new packages, active packages, new releases, active maintainers, and cumulative releases by year for CRAN.  Overall, these numbers clearly demonstrate a language that has experienced dramatic growth, with order-of-magnitude increases in packages, releases, and active maintainers.  Recent years have seen over 2,000 new packages, over 7,000 actively maintained packages, and over 1,600 active maintainers on CRAN.  Notable, however, is the peak and dip in growth of new packages and active maintainers from 2017 to 2019.  While the ecosystem has returned to growth in these metrics as of 2020, it is striking that CRAN experienced both a year-over-year decrease and an overall deceleration in growth given academic and industry interest in related fields.

\begin{table}[htbp]
    \centering
    \begin{tabular}{c|r|r|r|r|r}
    \hline
    \textbf{Year} &  \textbf{New Packages} & \textbf{Active Packages} & \textbf{New Releases} &  \textbf{Active Maintainers} & \textbf{Total Releases} \\
    \hline
    2000 &            43 &               43 &           104 &           19 &              43 \\
    2001 &            62 &               91 &           244 &           60 &             105 \\
    2002 &            52 &              125 &           341 &           47 &             157 \\
    2003 &            87 &              177 &           525 &           64 &             244 \\
    2004 &           124 &              279 &           867 &          101 &             368 \\
    2005 &           196 &              427 &          1,251 &          175 &             564 \\
    2006 &           250 &              555 &          1,622 &          202 &             814 \\
    2007 &           328 &              802 &          2,373 &          266 &            1,142 \\
    2008 &           386 &              913 &          2,456 &          319 &            1,528 \\
    2009 &           656 &             1,494 &          3,940 &          532 &            2,184 \\
    2010 &           721 &             1,747 &          4,533 &          583 &            2,905 \\
    2011 &           856 &             2,101 &          5,504 &          729 &            3,761 \\
    2012 &          1,039 &             2,782 &          6,015 &          938 &            4,800 \\
    2013 &          1,117 &             3,113 &          6,536 &         1,000 &            5,917 \\
    2014 &          1,312 &             3,413 &          6,772 &         1,088 &            7,229 \\
    2015 &          1,645 &             4,162 &          8,031 &         1,261 &            8,874 \\
    2016 &          2,143 &             4,921 &          9,710 &         1,478 &           11,017 \\
    2017 &          2,367 &             5,630 &         11,067 &         1,582 &           13,384 \\
    2018 &          2,087 &             6,432 &         12,230 &         1,497 &           15,471 \\
    2019 &          2,163 &             6,575 &         12,730 &         1,544 &           17,634 \\
    2020 &          2,294 &             7,530 &         14,630 &         1,637 &           19,928 \\
    \hline
    \end{tabular}
    \caption{Number of new packages, active packages, new releases, active maintainers and total releases on CRAN by year, 2020 is a partial year through December}
    \label{tab:cran_time_series}
\end{table}

Next, Table \ref{tab:bioconductor_time_series} details the same metrics for Bioconductor since Version 2.5.  As expected, Bioconductor's overall growth is much slower as a result of its focus on domain relevance and quality, and this is confirmed in its higher ratio of actively maintained packages to total releases than CRAN.  It is further unsurprising that Bioconductor does not exhibit the same kind of monotonic growth, as the committee's inclusion criteria and maintenance standards may even result in a \textit{reduction} of included packages in some cases.\footnote{Bioconductor's release page lists the number of packages available in each Bioconductor release version.  In some cases, these packages may be unchanged from prior Bioconductor releases in prior years.  As a result, while our numbers differ from Bioconductor's release numbers, ours reflect only new packages and releases in a calendar year and are comparable to other sources.}

\begin{table}[htbp]
    \centering
    \begin{tabular}{c|r|r|r|r|r}
    \hline
    \textbf{Year} &  \textbf{New Packages} & \textbf{Active Packages} & \textbf{New Releases} &  \textbf{Active Maintainers} & \textbf{Total Releases}\\
    \hline
    2009 &           106 &              107 &           878 &          104 &             175 \\
    2010 &            79 &              141 &           880 &           91 &             254 \\
    2011 &            72 &              161 &           875 &           79 &             326 \\
    2012 &            83 &              198 &          1,188 &          106 &             409 \\
    2013 &           124 &              238 &          1,194 &          123 &             533 \\
    2014 &           130 &              282 &          1,279 &          143 &             663 \\
    2015 &           169 &              353 &          1,704 &          172 &             832 \\
    2016 &           145 &              372 &          1,509 &          163 &             977 \\
    2017 &            69 &              312 &          1,110 &           79 &            1,046 \\
    2018 &           489 &              759 &          1,386 &          363 &            1,535 \\
    2019 &           333 &             1,021 &          2,202 &          289 &            1,868 \\
    2020 &           120 &             1,058 &          1,196 &          118 &            1,988 \\
    \hline
    \end{tabular}
    \caption{Number of new packages, active packages, new releases, active maintainers and total releases on Bioconductor by year}
    \label{tab:bioconductor_time_series}
\end{table}

Finally, in Table \ref{tab:github_time_series}, we show the last eight years of metrics from GitHub's releases.  GitHub exhibits the most rapid absolute increase out of all sources, likely reflecting the adoption of \textsc{devtools} by many developers of packages.  Recent years have seen approximately 1,000 new packages per year, approximately 1,500 actively maintained packages, and over 600 active maintainers on GitHub.  As discussed above in Section \ref{S:data}, some trivial amount of package activity is influenced by fork and pull request workflows, but this does not change our qualitative interpretations.

\begin{table}[htbp]
    \centering
    \begin{tabular}{c|r|r|r|r|r}
    \hline
    \textbf{Year} &  \textbf{New Packages} & \textbf{Active Packages} & \textbf{New Releases} &  \textbf{Active Maintainers} & \textbf{Total Releases}\\
    \hline
    2013 &            68 &               69 &           130 &           54 &              79 \\
    2014 &           382 &              407 &           827 &          284 &             461 \\
    2015 &           726 &              856 &          1,574 &          460 &            1,187 \\
    2016 &           999 &             1,257 &          2,355 &          660 &            2,186 \\
    2017 &          1,221 &             1,606 &          3,029 &          784 &            3,407 \\
    2018 &          1,323 &             1,810 &          3,572 &          884 &            4,730 \\
    2019 &           973 &             1,485 &          3,438 &          650 &            5,703 \\
    2020 &           947 &             1,549 &          4,279 &          624 &            6,650 \\
    \hline
    \end{tabular}
    \caption{Number of new packages, active packages, new releases, active maintainers and total releases on GitHub by year}
    \label{tab:github_time_series}
\end{table}

\begin{figure}[htbp]
    \centering
    \includegraphics[width=5in]{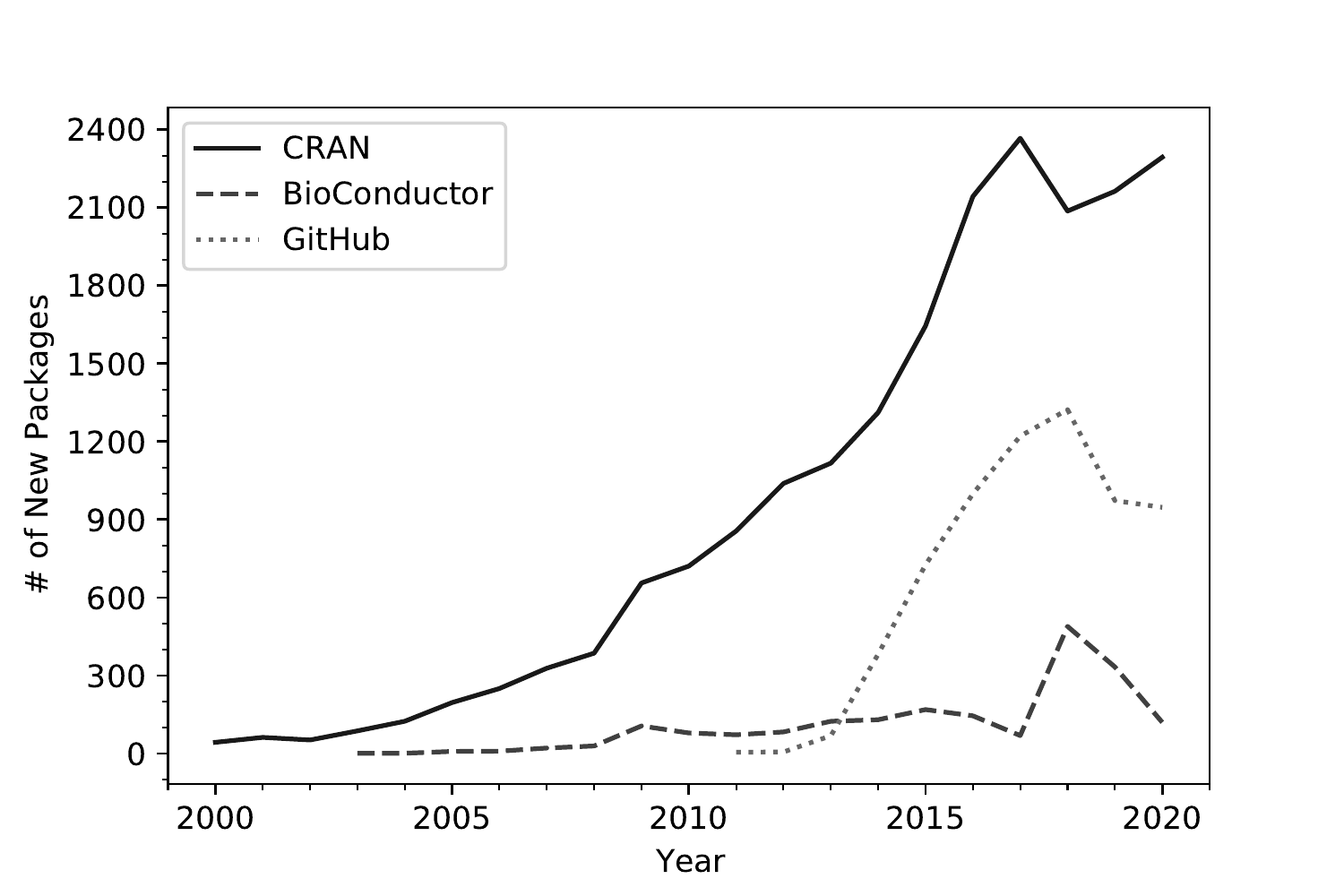}
    \caption{New packages by year across package sources.}
    \label{fig:packages_graph}
\end{figure}

\begin{figure}[htbp]
    \centering
    \includegraphics[width=5in]{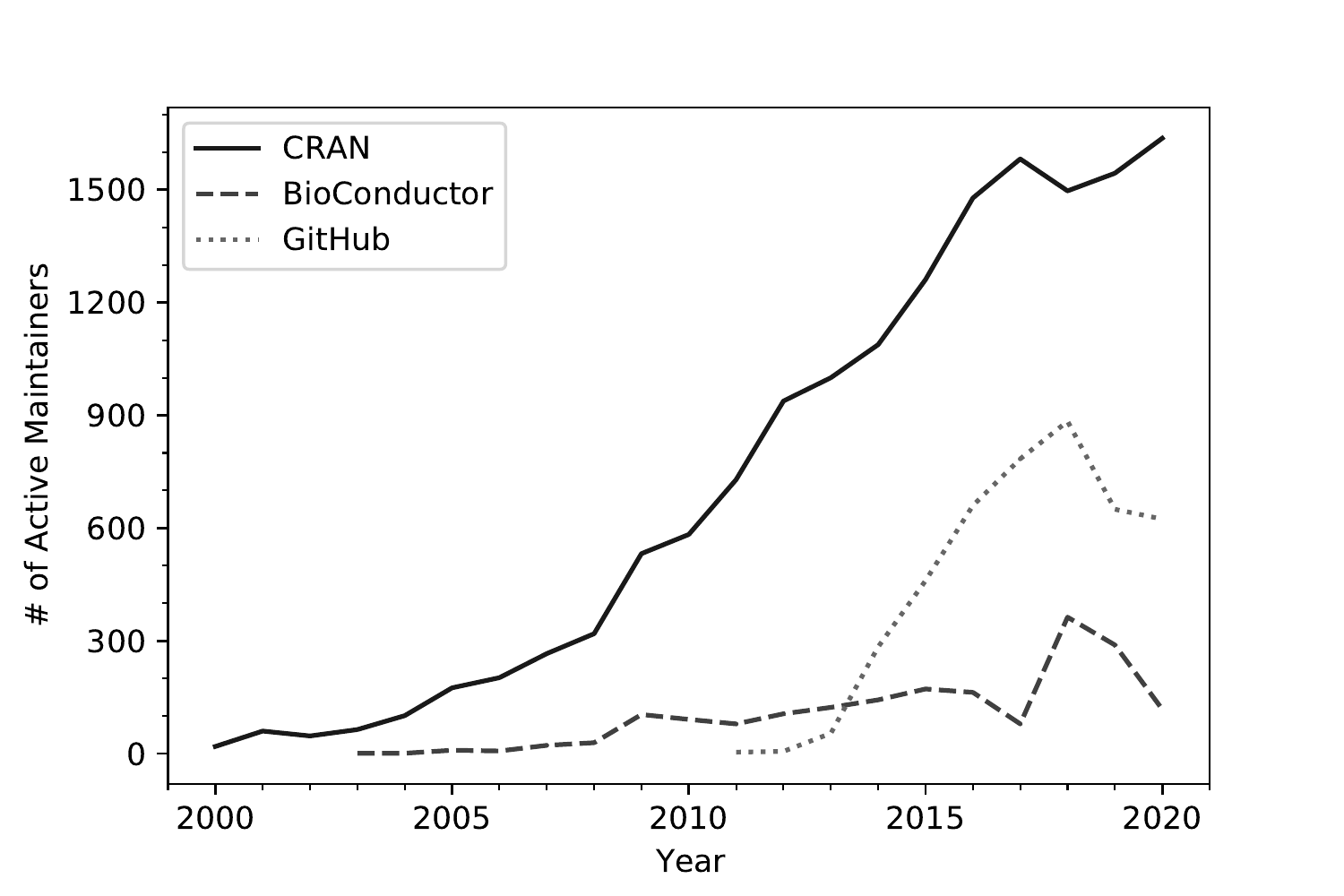}
    \caption{Active Maintainers by Year}
    \label{fig:maintainers_graph}
\end{figure}

When comparing the evolution of these three sources over time, there are a number of clear observations.  First, in alignment with our qualitative understanding of the sources, Bioconductor is the smallest but most actively maintained source of packages.  While the rate of new packages and releases does not increase monotonically over time, the packages that are distributed with Bioconductor receive updates and maintenance more regularly than packages on either GitHub or CRAN.  Second, while GitHub has rapidly established itself as an important forum for package development and distribution, it does not appear to be on a trajectory to replace CRAN.  Despite its open and approval-free release process, more packages have versions released annually on CRAN than GitHub still.  Figures \ref{fig:packages_graph} and \ref{fig:maintainers_graph} summarize these trends visually in line charts over time for both new packages and active maintainers.

Since these sources have existed for different durations and serve different needs, direct comparisons of raw counts can be misleading and mischaracterize the nature of the ecosystem.  Even growth rates may not be a useful metric for sources like Bioconductor, where one might expect a maturity and stability at a level of ``feature completion'' compared to an unconstrained scope.  Therefore, as described above, we synthesize these three sources into a single, cumulative time series that represents R's total ecosystem.  By viewing the sum of these sources together, we much more accurately capture the true activity and evolution of R over the last two decades.  We present the same metrics from Tables \ref{tab:cran_time_series}, \ref{tab:bioconductor_time_series}, and \ref{tab:github_time_series} for the total ecosystem in Table \ref{tab:total_time_series}.  

\begin{table}[htbp]
    \centering
    \begin{tabular}{c|r|r|r|r|r}
    \hline
    \textbf{Year} &  \textbf{New Packages} & \textbf{Active Packages} & \textbf{New Releases} &  \textbf{Active Maintainers} & \textbf{Total Releases} \\
    \hline
    2000 &           43 &              43 &          104 &          19 &             43 \\
    2001 &           62 &              91 &          244 &          60 &            105 \\
    2002 &           52 &             125 &          341 &          47 &            157 \\
    2003 &           88 &             178 &          546 &          65 &            245 \\
    2004 &          125 &             280 &          871 &         102 &            370 \\
    2005 &          202 &             434 &        1,393 &         182 &            572 \\
    2006 &          259 &             564 &        1,741 &         209 &            831 \\
    2007 &          347 &             821 &        2,667 &         285 &          1,178 \\
    2008 &          415 &             942 &        2,771 &         347 &          1,593 \\
    2009 &          759 &           1,597 &        4,802 &         625 &          2,352 \\
    2010 &          799 &           1,877 &        5,387 &         663 &          3,151 \\
    2011 &          930 &           2,256 &        6,376 &         801 &          4,081 \\
    2012 &        1,123 &           2,983 &        7,208 &       1,040 &          5,204 \\
    2013 &        1,290 &           3,393 &        7,831 &       1,153 &          6,494 \\
    2014 &        1,761 &           3,984 &        8,684 &       1,474 &          8,255 \\
    2015 &        2,423 &           5,120 &       10,929 &       1,812 &         10,678 \\
    2016 &        3,105 &           6,159 &       12,915 &       2,206 &         13,783 \\
    2017 &        3,456 &           7,022 &       14,341 &       2,354 &         17,239 \\
    2018 &        3,748 &           8,452 &       16,285 &       2,633 &         20,987 \\
    2019 &        3,320 &           8,504 &       17,220 &       2,390 &         24,307 \\
    2020 &        3,205 &           9,462 &       18,637 &       2,274 &         27,512 \\
    \hline
    \end{tabular}
    \caption{Number of new packages, active packages, new releases, active maintainers and total releases across repositories by year}
    \label{tab:total_time_series}
\end{table}

Table \ref{tab:total_time_series} documents an impressive two decades for the R ecosystem.  In its first five years, R's package ecosystem slowly grew to approximately 250 active packages and 100 maintainers.   In the next five years, these numbers grew more than five-fold to over 1,700 active packages and nearly 600 maintainers.  Between 2010 and 2015, these numbers more than doubled to over 4,000 active packages and over 1,200 maintainers.  Finally, in the most recent five years, this growth has slowed but still resulted in over 7,500 active packages and 1,600 maintainers.

To better summarize this growth across R and within each source, we calculate the compound annual growth rate (CAGR) for each source and the collective ecosystem.  These results are presented for new packages, active packages, new releases, and active maintainers below in Table \ref{tab:ecosystem_cagr}.  These calculations confirm our observations above and provide a convenient summarization of R's two decades of growth.  On average, R has sustained a total growth rate of approximately 29\% for active packages, and 28\% for new releases, and 26\% for active maintainers.  For comparison with Python, we document a CAGR of 47\% for active packages and 39\% for maintainers within PyPI in \cite{bommarito2019empirical}.  While this cumulative growth has been impressive, it is worth noting that recent years indicate a potential peak or plateau in R's growth.  Since 2017-2018, the number of new packages and active maintainers on both GitHub and CRAN has remained relatively constant or decreased.  We will continue to monitor this trend through 2021 to examine whether the R ecosystem may be stable or actually decreasing at this point.

\begin{table}[htbp]
    \centering
    \begin{tabular}{r|r|r|r|r}
    \hline
    \textbf{Measure} &         \textbf{CRAN} & \textbf{Bioconductor} & \textbf{GitHub}  & \textbf{Total Ecosystem}\\ 
    \hline
    New Packages       & 20.85\% &  1.04\% & 38.99\% &   22.49\% \\
    Active Packages    & 27.89\% & 21.04\% & 47.54\% &   29.29\% \\
    New Releases       & 26.56\% &  2.61\% & 54.77\% &   28.03\% \\
    Active Maintainers & 23.64\% &  1.06\% & 35.78\% &   25.61\% \\
    Total Releases     & 33.95\% & 22.45\% & 74.04\% &   35.82\% \\
    \hline
    \end{tabular}
    \caption{Compound Annual Growth Rate (CAGR) for new packages, new releases and active maintainers on CRAN.  CAGR calculation starts at 2000 for CRAN, 2009 for BIOC, 2013 for GitHub, and 2000 for combined.}
    \label{tab:ecosystem_cagr}
\end{table}

\subsection{Packages and Releases}
\label{s:results_packages}
In order to better understand package dynamics generally and within each source, we next examine per-package distributions and statistics.  First, we calculate summary statistics for the distribution of releases per package in Table \ref{tab:ecosystem_package_release}.  This distribution helps us understand how packages are, or are not, maintained by each source and overall.  The difference between the GitHub, CRAN and Bioconductor is immediately apparent from this table.  GitHub is the most right-skewed of these sources, as expected, with a median of 1.0 release per package and a mean of 2.9.  Many repositories in GitHub are forks or single releases that see little maintenance or are even entirely abandoned. There are repositories that see regular updates, such as popular projects developed by the R core team, academic groups, or companies.  CRAN shows some right-skewness as well, though much less, as its median releases per package is 3.0 and mean is 5.8.  Bioconductor, by contrast, has both the highest mean releases per package and an almost identical median, demonstrating the quality control and support provided through its curation.  On average across the ecosystem, most R packages have had 3.0 or fewer releases with an average of approximately 5.0.

\begin{table}[htbp]
    \centering
    \begin{tabular}{r|r|r|r|r}
    \toprule
    \hline
    \textbf{Statistic} &          \textbf{CRAN} &         \textbf{Bioconductor} & \textbf{GitHub} & \textbf{Across Repositories}\\
    \hline
    \midrule
    Mean  &      5.75 &    10.44 &     2.89 &      5.60 \\
    Standard Deviation   &      8.76 &     6.40 &     5.54 &      8.51 \\
    Minimum   &      1.00 &     1.00 &     1.00 &      1.00 \\
    25th Percentile   &      1.00 &     5.00 &     1.00 &      1.00 \\
    50th Percentile   &      3.00 &    10.00 &     1.00 &      3.00 \\
    75th Percentile   &      7.00 &    15.00 &     3.00 &      7.00 \\
    Maximum   &    200.00 &    21.00 &   201.00 &    250.00 \\
    \hline
    \end{tabular}
    \caption{Descriptive statistics for distribution of number of releases per package}
    \label{tab:ecosystem_package_release}
\end{table}

We next examine the distribution of duration between releases or inter-release timing for packages.  Table \ref{tab:ecosystem_inter_release} below presents the mean, standard deviation, min, max, and quartiles for this distribution by source and across the entire ecosystem.  Packages that have a lower duration between releases are likely higher velocity and/or younger packages, whereas packages with more time between releases may be either mature or lower velocity.  As Bioconductor has a defined semi-annual release calendar, its timing is generally less varied.  For CRAN, the mean and median inter-release durations are 213 and 96 days.  For GitHub, these mean and median durations are 100 and 37 days, respectively.  Unsurprisingly, packages that are active and released on GitHub  demonstrate a higher velocity than those released through CRAN, likely as a result of decreased submission and approval process friction.  On average across the ecosystem, packages release approximately six months apart, though there is substantial variation.

\begin{table}[htbp]
    \centering
    \begin{tabular}{r|r|r|r|r}
    \hline
    \textbf{Measure} &         \textbf{CRAN} & \textbf{Bioconductor} & \textbf{GitHub} & \textbf{Across Repositories}\\
    \hline
    Mean  &  213 days  &  283 days   &  100 days  &  178 days  \\
    Standard Deviation   &  331 days  &  343 days   &  172 days   &  312 days  \\
    Minimum   &              1 days &              1 days  &              1 days &              0 days \\
    25th Percentile   &             33 days &            161 days &              8 days  &             11 days  \\
    50th Percentile   &             96 days  &            181 days   &             37 days  &             68 days \\
    75th Percentile   &            247 days  &            286 days &            114 days  &            200 days  \\
    Maximum   &           5,236 days  &           4,658 days &           2,182 days  &           5,236 days  \\
    \hline
    \end{tabular}
    \caption{Descriptive statistics for inter-release duration distribution}
    \label{tab:ecosystem_inter_release}
\end{table}

To better understand the right tail of this distribution - packages with many releases - we examine the top 20 packages by release count.  These packages are listed below in Table \ref{tab:top_package_release}.  In general, most of these packages have long histories on CRAN, e.g., \textsc{Matrix}, or are built by active teams or companies on GitHub, e.g., \textsc{canvasXpress}.  Many of these packages are even authored and maintained by the R-core development team and its members, such as \textsc{nlme}, \textsc{lattice}, \textsc{Matrix}, and \textsc{mgcv}.

\begin{table}[htbp]
    \centering
    \begin{tabular}{r|r}
    \hline
    \textbf{Name} &  \textbf{Count} \\
    \hline
    spatstat      &       250 \\
    PortalData    &       201 \\
    canvasXpress  &       199 \\
    Matrix        &       198 \\
    pomp          &       162 \\
    mgcv          &       155 \\
    nlme          &       136 \\
    RcppArmadillo &       136 \\
    lattice       &       135 \\
    rgdal         &       134 \\
    caret         &       133 \\
    spdep         &       129 \\
    plotrix       &       129 \\
    sp            &       125 \\
    Rcmdr         &       121 \\
    XML           &       121 \\
    gstat         &       118 \\
    arm           &       117 \\
    lme4          &       117 \\
    foreign       &       115 \\
    \hline
    \end{tabular}
    \caption{Top 20 packages by number of releases across repositories}
    \label{tab:top_package_release}
\end{table}

\subsection{Contributors}
\label{s:results_contributors}
Releases and packages do not, of course, grow on trees.  Contributors of a wide variety create and sustain these works, and, in order to truly understand the ecosystem, we must examine dynamics related to these individuals and organizations.  Maintainer data, the most common of these contributor types, is provided in numerous forms within the R ecosystem.  In all packages containing DESCRIPTION files, an \textsc{Authors\@R} field can contain a list of one or more individuals or organizations, including details about the role that they play for that package or release.  In addition, packages published to CRAN must include a Maintainer field for an author who provides an email address.  Table \ref{tab:total_package_maintainer} below provides a summary of the distribution of packages and releases per maintainer.   As is the case in many records of human creators, the majority of maintainers have a single work while a small number of maintainers is responsible for many packages; the median number of packages and releases per maintainer is 1 and 3, respectively.  Even at the 75th percentile, these medians increase to just 2 packages and 8 releases.  

\begin{table}[htbp]
    \centering
    \begin{tabular}{r|r|r}
    \toprule
    \hline
    \textbf{Statistic} &  \textbf{Packages per Maintainer} &  \textbf{Releases per Maintainer} \\
    \hline
    \midrule
    Mean  &                    1.81 &                    8.90 \\
    Standard Deviation   &                    2.56 &                   24.62 \\
    Minimum   &                    1.00 &                    1.00 \\
    25th Percentile   &                    1.00 &                    1.00 \\
    50th Percentile   &                    1.00 &                    3.00 \\
    75th Percentile   &                    2.00 &                    8.00 \\
    Maximum   &                   99.00 &                  791.00 \\
    \bottomrule
    \hline
    \end{tabular}
    \caption{Descriptive statistics for distribution of packages and releases per maintainer across repositories}
    \label{tab:total_package_maintainer}
\end{table}

In Table \ref{tab:top_maintainers}, we examine the right tail of the distribution in Table \ref{tab:total_package_maintainer} by listing the top 20 maintainers by package counts. Many of these authors should come as  no surprise, as Hadley and Dirk might as well be listed as essential personnel in R's ecosystem.  Some of these authors, however, are less well-known to the community, such as Jia Zhong, who is responsible for IBM's \textsc{IBMPredictiveAnalytics} GitHub repository.  In some cases, these authors are actually organizations or their automation, such as the Bioconductor Package Maintainer metadata author description.

\begin{table}[htbp]
    \centering
    \begin{tabular}{r|r}
    \hline
    \textbf{Name} &  \textbf{Count} \\
    \hline
    Scott Chamberlain                &               99 \\
    Hadley Wickham                   &               78 \\
    Richel Bilderbeek                &               69 \\
    Dirk Eddelbuettel                &               66 \\
    Jia Zhong Wu                     &               64 \\
    Gábor Csárdi                     &               57 \\
    Bioconductor Package Maintainer  &               56 \\
    Jeroen Ooms                      &               54 \\
    Thomas J. Leeper                 &               38 \\
    Kurt Hornik                      &               37 \\
    Scott Chamberlain                &               36 \\
    Max Kuhn                         &               35 \\
    Bob Rudis                        &               34 \\
    Robin K. S. Hankin               &               34 \\
    Hadley Wickham                   &               34 \\
    Martin Maechler                  &               34 \\
    Gabor Csardi                     &               34 \\
    Kartikeya Bolar                  &               33 \\
    Henrik Bengtsson                 &               32 \\
    Jan Wijffels                     &               31 \\
    \hline
    \end{tabular}
    \caption{Top 20 maintainers by number of packages maintained across repositories}
    \label{tab:top_maintainers}
\end{table}

Not all packages are equal in effort, of course.  Packages vary both in terms of their size, composition, and complexity; some might contain primarily data instead of source code, and others might contain a small number of very complex functions.  In order to provide a view into the distribution of effort in maintaining packages, we present the top maintainers by number of lines of code (LOC) in Table \ref{tab:top_maintainer_loc}.  These LOC calculations include only files such as R, Fortran, C, and C++ source files, excluding any lines or bytes from data, auto-generated documentation, or other package contents. Unsurprisingly, a small number of authors are responsible for an outsized percentage of R code; the top 20 maintainers alone support over 15\% of all lines of code in the R ecosystem.  

\begin{table}[htbp]
    \centering
    \begin{tabular}{r|r|r}
    \toprule
    \hline
    \textbf{Maintainer} &       \textbf{KLOC} &     \textbf{Percent of total KLOC}\\
    \hline
    \midrule
    Michael Lawrence     & 15,727.78 & 1.64\% \\
    Adrian Baddeley      & 10,878.97 & 1.13\% \\
    Marek Gagolewski     & 10,326.76 & 1.08\% \\
    weizhouUMICH         & 10,038.94 & 1.05\% \\
    Henrik Bengtsson     &  9,527.69 & 0.99\% \\
    Adrian Baddeley      &  8,299.77 & 0.86\% \\
    Joshua N. Pritikin   &  8,230.79 & 0.86\% \\
    Gabor Csardi         &  7,367.14 & 0.77\% \\
    Kurt Hornik          &  7,193.62 & 0.75\% \\
    Martin Maechler      &  6,823.04 & 0.71\% \\
    Stan buildbot        &  6,706.61 & 0.70\% \\
    Wei-Chen Chen        &  6,209.87 & 0.65\% \\
    Douglas Bates        &  6,035.10 & 0.63\% \\
    Jeffrey S. Racine    &  5,074.88 & 0.53\% \\
    Roger Bivand         &  5,022.16 & 0.52\% \\
    Edzer Pebesma        &  4,801.93 & 0.50\% \\
    Vladislav Kim        &  4,372.42 & 0.46\% \\
    Doug and Martin      &  4,367.44 & 0.46\% \\
    Virginie Rondeau     &  3,978.60 & 0.41\% \\
    Alexander Robitzsch  &  3,970.17 & 0.41\% \\
    \bottomrule
    \end{tabular}
    \caption{Top 20 maintainers by KLOC}
    \label{tab:top_maintainer_loc}
\end{table}

Conversely, many packages also have more than just one contributor, whether that contributor is an individual or organization.  Luckily, many releases explicitly include additional contributors in their \textsc{DESCRIPTION} metadata. Table \ref{tab:total_package_contributor} details the descriptive statistics for the distribution of contributors per package across repositories. The mean and median number of listed contributors per package is 2.5 and 2.0, respectively, increasing to 3.0 at the 75th percentile.  The right tail of this distribution is marked by packages that distribute collections of data from multiple sources.  For example, the package with 127 contributors is the \textsc{rcorpora} package, which includes a ``collection of small text corpora of interesting data.''

\begin{table}[htbp]
    \centering
    \begin{tabular}{r|r|}
    \toprule
    \hline
    \textbf{Statistic} &  \textbf{Contributors per Package}\\
    \hline
    \midrule
    Mean  &      2.46 \\
    Standard Deviation   &      3.05 \\
    Minimum   &      1.00 \\
    25th Percentile   &      1.00 \\
    50th Percentile   &      2.00 \\
    75th Percentile   &      3.00 \\
    Maximum   &    127.00 \\
    \bottomrule
    \hline
    \end{tabular}
    \caption{Descriptive statistics for distribution of contributors per package across repositories}
    \label{tab:total_package_contributor}
\end{table}

Within R \textsc{DESCRIPTION}, packages may also specify the \textsc{role} of an author or contributor to a package.  These roles are based on the Library of Congress's MARC Code List for Relators \cite{loc2021}, though as they are manually entered without validation, some data quality issues occur.  Table \ref{tab:top_contributor_tags} presents counts of the most common three-letter role designations.  Some of these tags clearly reflect R's usage in academia, e.g., the fifth most popular tag is ``Thesis Advisor.''  Many other tags highlight the unreliability of such metadata, either due to common misspellings or humorous self-reported roles like woodcutter.  As an example, all contributors to the \textsc{rcorpora} discussed above are listed without a role tag.

\begin{table}[htbp]
    \centering
    \begin{tabular}{r|r}
    \hline
    \textbf{Tag} &  \textbf{Count} \\
    \hline
    Untagged         &  184,996 \\
    Author           &   92,357 \\
    Creator          &   49,071 \\
    Contributor      &   48,824 \\
    Copyright Holder &   17,457 \\
    Thesis Advisor   &    1,510 \\
    Translator       &     793 \\
    Funder           &     725 \\
    Data Contributor &     725 \\
    Contractor       &     334 \\
    \hline
    \end{tabular}
    \caption{Top 10 package contributor tags across repositories}
    \label{tab:top_contributor_tags}
\end{table}

In general, our analysis suggests that the R ecosystem leans heavily on a small number of contributing individuals and organizations, but that there are many potential sources of confusion from a data quality perspective.  The community would likely benefit from a concerted effort to solicit maintainer assistance and to improve validation for metadata through \textsc{R CMD} functionality.  Such improvements in data quality, especially as they relate to copyright holders and licensing, may be critical for the further adoption of R in industry.

\subsection{Dependency Network}
\label{s:results_dependency}
Based on our author data above, just 0.1\% of maintainers are responsible for 3.4\% of packages.  However, not all packages are equally important in the ecosystem, as some packages may be relied on much more frequently as a dependency than others. \cite{MORACANTALLOPS2020110744} has already documented the existence of a right-skewed distribution for package in-degree in a similar but smaller, non-longitudinal sample of CRAN.  The interested reader is referred to their research for more detail on the structure of R package networks, such as macro- and mesoscopic community analysis.  

In Table \ref{tab:in-dependent_summary}, we begin by examining the in-degree distribution for packages in our sample.  Conceptually, packages that have higher in-degree are typically more ``important'' or popular in the ecosystem, just as food web analyses can reveal important organisms in real ecological systems.  As our sample is longitudinal and spans multiple package sources, some of which do not automatically connect, it is important to note that our results do not characterize the \textit{current} CRAN network alone.  Further, just as in ecological systems, some organisms may have critical roles by virtue of network structure and their position, which is not captured by in-degree alone.  The interested reader is again directed to \cite{MORACANTALLOPS2020110744} \cite{mora2020evolution} for more on such structural analysis.  We note, however, that both prior literate and our research do not reflect the use of libraries in ad-hoc statistical analysis or ``data science'' scripts, where users rarely formally save or package the sequence of executed statements. 

Within our sample, the mean and median package in-degrees are 1.93 and 0.0, confirming that the distribution is right-skewed.  In fact, even the 75th percentile is 0.0, i.e., more than three out of four packages have never been imported as a dependency.  This finding is not surprising when one considers that many R packages are specialized statistical methods or point solutions with narrow application or domain; therefore, while many researchers may use these packages to perform ad-hoc statistical analyses, there is little reason to ``build on top of'' these packages.

\begin{table}[htbp]
    \centering
    \begin{tabular}{r|r}
    \toprule
    \hline
    \textbf{Statistic} &  \textbf{Dependency}\\
    \hline
    \midrule
    Mean  &      1.93 \\
    Standard Deviation   &     26.01 \\
    Minimum   &      0.00 \\
    25th Percentile   &      0.00 \\
    50th Percentile   &      0.00 \\
    75th Percentile   &      0.00 \\
    Maximum   &  1,845.00 \\
    \bottomrule
    \hline
    \end{tabular}
    \caption{Descriptive statistics for distribution of in-dependencies per package across repositories}
    \label{tab:in-dependent_summary}
\end{table}

Next, we examine the out-degree distribution for packages in our sample in Table \ref{tab:out-dependent_summary}.  The mean and median out-degree per package are 1.93 and 1.0, less skewed than in-degree.  Further, while the 75th percentile out-degree, 3.0, is higher than the corresponding in-degree quartile, the maximum out-degree is two orders of magnitude smaller for out-degree than in-degree. Taken together, these in- and out-degree distributions confirm the intuition that R contains a small number of very important packages and a large number of loosely-connected or singleton packages.

\begin{table}[htbp]
    \centering
    \begin{tabular}{r|r}
    \toprule
    \hline
    \textbf{Statistic} &  \textbf{Dependency}\\
    \hline
    \midrule
    Mean  &      1.93 \\
    Standard Deviation   &      3.32 \\
    Minimum   &      0.00 \\
    25th Percentile   &      0.00 \\
    50th Percentile   &      1.00 \\
    75th Percentile   &      3.00 \\
    Maximum   &     59.00 \\
    \bottomrule
    \hline
    \end{tabular}
    \caption{Descriptive statistics for distribution of out-dependencies per package across repositories}
    \label{tab:out-dependent_summary}
\end{table}

As shown in Table \ref{tab:top_dependencies}, packages at the extreme right tail such as \textsc{dplyr} have over 1,000 unique package dependents.  The table lists the twenty most depended-on, i.e., highest in-degree, packages. There are a number of immediate observations. Many of the most popular packages are intended to simplify syntax, provide foundational data structures, or make pre-processing data easier.  Rcpp is relied on by many CRAN packages that provide acceleration through \textsc{C++} or vendor third-party \textsc{C++} libraries.  When in-degree is calculated based on release instead of package, the ranks are slightly different but the distribution is even more extreme; for example, \textsc{dplyr} and \textsc{ggplot2} are a dependent each for 7,701 and 8,194 separate releases, respectively.  Overall, the extreme dependence is notable compared to other languages; the top five packages are imported by nearly one in every four packages. 

\begin{table}[htbp]
    \centering
    \begin{tabular}{r|r}
    \hline
    \textbf{Name} &  \textbf{Count} \\
    \hline
    dplyr         &             1,845 \\
    ggplot2       &             1,798 \\
    MASS          &             1,178 \\
    Rcpp          &             1,026 \\
    magrittr      &              907 \\
    data.table    &              860 \\
    Matrix        &              780 \\
    jsonlite      &              665 \\
    httr          &              639 \\
    Biobase       &              562 \\
    foreach       &              550 \\
    plyr          &              480 \\
    GenomicRanges &              463 \\
    IRanges       &              463 \\
    BiocGenerics  &              453 \\
    S4Vectors     &              452 \\
    igraph        &              448 \\
    doParallel    &              432 \\
    purrr         &              409 \\
    lattice       &              407 \\
    \hline
    \end{tabular}
    \caption{Top 20 dependencies by number of packages across repositories}
    \label{tab:top_dependencies}
\end{table}

Strikingly, a number of the most depended-on packages are maintained by the same contributors.\footnote{These calculations count unique dependency per package; multiple releases do not weight or multiply edges or degree, and even if dependencies are removed in later releases, we still record such imports.}  In order to understand just how much concentration there is, we examine how in-degree is distributed across maintainers in the ecosystem.  Table \ref{tab:top_maintainer_indegree} shows the top 20 maintainers by this measure, and the results confirm our intuitions again.  Hadley, for example, is the listed maintainer responsible for packages that are imported by over 4,000 other packages.  When one considers that a number of these contributors are also employed by RStudio, the concentration of support effort becomes even more extreme.  In total, the top 10 maintainers are responsible for nearly 16,000 unique package dependencies; this means that more than one out of every two packages relies on these ten maintainers.  Such extreme reliance on a small number of packages and persons has implications for ecosystem sustainability and potential supply chain attacks.

\begin{table}[htbp]
    \centering
    \begin{tabular}{r|r}
    \toprule
    \hline
    \textbf{Maintainer} & \textbf{Number of Packages Dependent} \\
    \midrule
    \hline
    Hadley Wickham                         &                4,321 \\
    Thomas Lin Pedersen                    &                1,824 \\
    Brian Ripley                           &                1,633 \\
    Bioconductor Package Maintainer        &                1,390 \\
    Dirk Eddelbuettel                      &                1,344 \\
    Lionel Henry                           &                1,218 \\
    Martin Maechler                        &                1,195 \\
    Dirk Eddelbuettel and Romain Francois  &                1,028 \\
    Douglas Bates                          &                1,015 \\
    Jeroen Ooms                            &                  928 \\
    Stefan Milton Bache                    &                  907 \\
    Doug and Martin                        &                  901 \\
    Gábor Csárdi                           &                  887 \\
    M Dowle and T Short                    &                  860 \\
    Developers                             &                  860 \\
    Gabor Csardi                           &                  720 \\
    Biocore Team c/o BioC user list        &                  687 \\
    Hong Ooi                               &                  644 \\
    Jim Hester                             &                  642 \\
    Rich Calaway                           &                  633 \\
    \bottomrule
    \end{tabular}
    \caption{Top 20 maintainers by number of packages dependent on their maintained packages}
    \label{tab:top_maintainer_indegree}
\end{table}

\begin{figure}[htbp]
    \centering
    \includegraphics{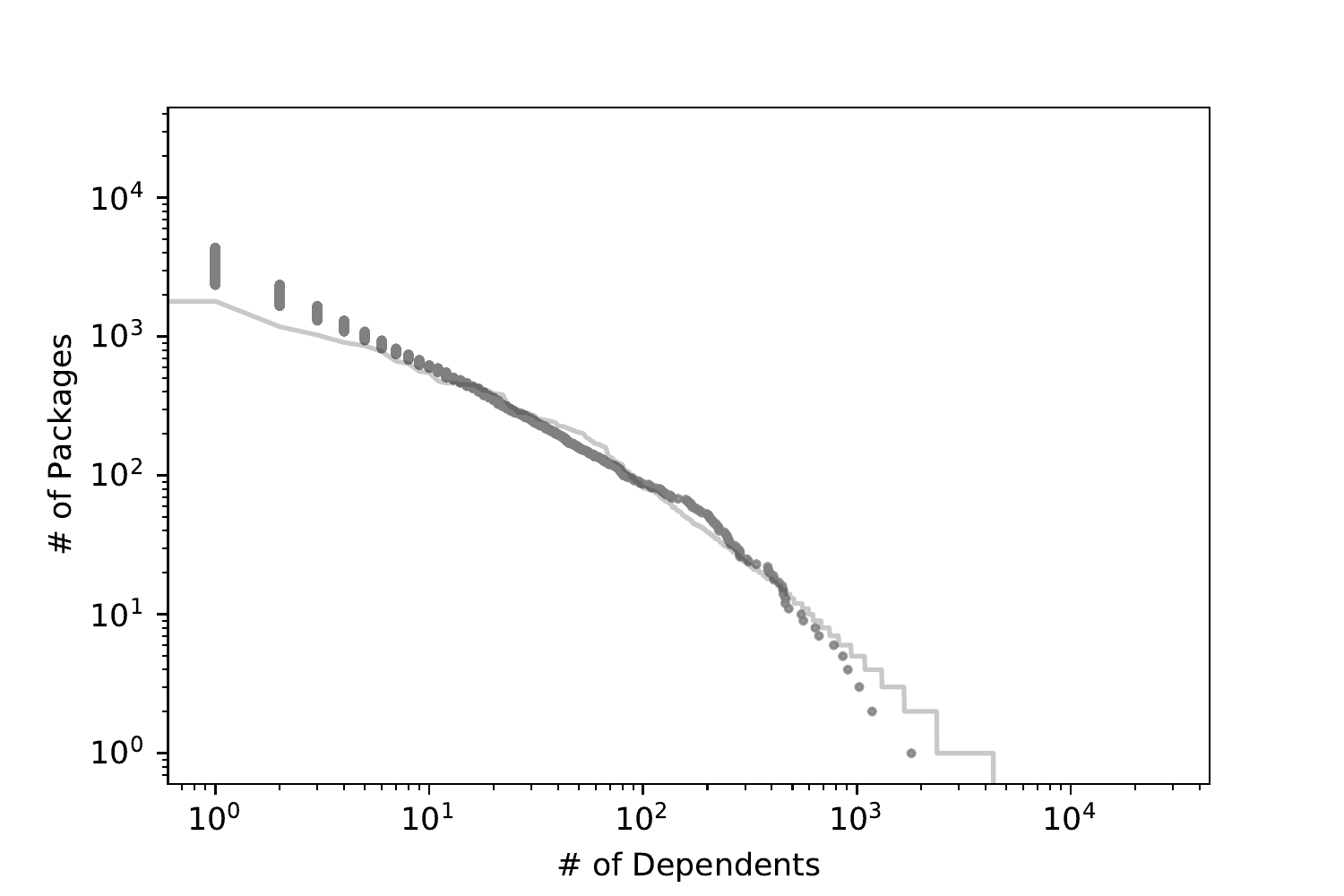}
    \caption{Number of Dependents by Package}
    \label{fig:dependencies_graph}
\end{figure}

For readers interested in how our in- and out-degree distributions compare to other right-skewed distributions or the prior work mentioned above, we have calculated log-log plots of degree distributions in Figure \ref{fig:dependencies_graph}.  As noted above, our results are longitudinal and span multiple package sources; therefore, they are larger and conceptually different from static, single-source analyses of dependency networks. 

\subsection{Licensing}
Licensing is a key factor for the success of open-source software ecosystems.  Depending on the composition and usage of software within the ecosystem, there is risk of both free-riding if licenses are too permissive and under-investment if licenses are too restrictive.  The R Project itself, including the R interpreter and core libraries, is currently licensed under the GPL Version 2.  While a complete discussion of licensing factors and analysis of R licensing is outside the scope of this paper, we present a summary distribution of license metadata in Table \ref{tab:top_licenses}.  Unfortunately, the licensing metadata situation for R is complicated by a mixture of license restrictions and poor metadata management.  For instance, there are nearly two thousand releases that have no listed license in their metadata.  Many other metadata fields specify a license identifier, but also indicate that a separate \textsc{LICENSE} file, which may contradict the metadata field, is also present.  

Excluding omitted license metadata, the GPL family licenses such as GPL Version 2 and 3 outnumber all other licenses like MIT, BSD, or Apache family licenses.  Compared with other open-source ecosystems, this is a striking finding; from a proportion perspective, GPL is an order of magnitude more common in R than in Python, for example.  This preference may be due to the original R core team's preference for GPL licensing.  In future research, we will present more detailed analysis of licensing and compliance trends within R packages and source files, documenting trends such as re-licensing, ``mixed'' or multi-licensing, data sub-``licensing'' or restrictions, and contradictory license indications.

\begin{table}[htbp]
    \centering
    \begin{tabular}{r|r|r}
    \hline
    \textbf{License} &  \textbf{Count} & \textbf{Percent}\\
    \hline
    No License in Metadata    & 1,707.0 &  5.94\% \\
    GPL 2                     & 9,583.0 & 33.37\% \\
    GPL 3                     & 5,856.0 & 20.39\% \\
    MIT + file LICENSE        & 3,286.0 & 11.44\% \\
    GPL                       &   826.0 &  2.88\% \\
    Artistic-2.0              &   597.0 &  2.08\% \\
    GPL 3 | file LICENSE      &   275.0 &  0.96\% \\
    CC0                       &   260.0 &  0.91\% \\
    LGPL-3                    &   213.0 &  0.74\% \\
    file LICENSE              &   182.0 &  0.63\% \\
    MIT                       &   161.0 &  0.56\% \\
    BSD 3 + file LICENSE      &   158.0 &  0.55\% \\
    GPL 2 | file LICENSE      &   152.0 &  0.53\% \\
    Apache License 2.0        &   152.0 &  0.53\% \\
    BSD 2 + file LICENSE      &   150.0 &  0.52\% \\
    GPL ($>=$ 2.0)              &   141.0 &  0.49\% \\
    AGPL-3                    &   140.0 &  0.49\% \\
    LGPL                      &   134.0 &  0.47\% \\
    What license is it under? &   102.0 &  0.36\% \\
    GPL 3 + file LICENSE      &    74.0 &  0.26\% \\
    Unlimited                 &    73.0 &  0.25\% \\
    \hline
    \end{tabular}
    \caption{Top 20 Licenses (metadata) by Number of Releases}
    \label{tab:top_licenses}
\end{table}

\subsection{Source Analysis}
Unlike most extant research, our data set and analysis also includes full file-level records and file-level detail data.  For example, we present in Table \ref{tab:loc_summary} a summary of the average number of lines of code per file and per release for R, Fortran, C, and C++ files across all sources.  Many packages in the R ecosystem contain not just R source files, but also Fortran, C, or C++ sources, either for acceleration or to vendor from third party packages.  While Fortran, C, and C++ files are, on average, longer than R source files - an unsurprising finding given the languages - there are over twice as many lines of R code across all sources as there are Fortran, C, and C++ lines together.

\begin{table}[htbp]
    \centering
    \begin{tabular}{r|r|r|r|r}
    \toprule
    \hline
    \textbf{Statistic} &     \textbf{R} &     \textbf{C} &  \textbf{C++} & \textbf{Fortran}\\
    \midrule
    \hline
    Average Per File    &   157.31 &   527.21 &     308.30 &     528.44 \\
    Average Per Release & 4,208.14 & 1,118.92 &     614.00 &     223.84 \\
    \bottomrule
    \hline
    \end{tabular}
    \caption{Summary of LOC across sources by source file type}
    \label{tab:loc_summary}
\end{table}

Table \ref{tab:loc_time_series} presents trends in R, Fortran, C, and C++ source KLOC over time, detailing how many lines of code by language are available in the latest release for all packages as of each year.  In the first decade of R packages, Fortran played an important role; for example, in 2005, there were 5 LOC of R for every one of Fortran, and in 2010, there were 7 LOC of R for every one of Fortran.  By 2015, there were more than 10 LOC of R for every one of Fortran, and by 2020, there are nearly 30 LOC of R for every one of Fortran.  C has similarly fallen from near-parity with R to approximately 1:7 in 2020.  Prior to the release of \textsc{Rcpp} in late 2008, there were very few lines of C++; however, after the introduction of \textsc{Rcpp}, C++ has grown to nearly the same proportion as C in recent years.  When understanding and auditing R sources, it is clearly critical that the licenses and security of these C, C++, and Fortran sources are considered in addition to R alone.

\begin{table}[htbp]
    \centering
    \begin{tabular}{r|r|r|r|r}
    \hline
    \toprule
    \textbf{Year} &     \textbf{R} &     \textbf{C} &  \textbf{C++} & \textbf{Fortran}\\
    \hline
    \midrule
    2000 &    124.38 &     64.50 &      0.00 &    10.11 \\
    2001 &    237.45 &    128.45 &      2.16 &    55.06 \\
    2002 &    344.06 &    185.97 &     12.02 &    57.59 \\
    2003 &    562.87 &    298.27 &     31.98 &    93.66 \\
    2004 &    854.26 &    580.74 &     53.25 &   124.70 \\
    2005 &  1,315.63 &    670.51 &     65.52 &   228.96 \\
    2006 &  1,919.82 &  1,025.85 &    122.51 &   333.65 \\
    2007 &  2,949.41 &  1,545.85 &    208.96 &   524.56 \\
    2008 &  3,823.46 &  1,917.84 &    249.68 &   616.18 \\
    2009 &  5,477.38 &  2,638.28 &    813.07 &   886.45 \\
    2010 &  7,415.01 &  3,181.40 &  1,113.12 & 1,065.02 \\
    2011 &  8,814.38 &  3,794.45 &  1,328.33 & 1,221.56 \\
    2012 & 11,247.18 &  4,898.48 &  1,870.03 & 1,492.94 \\
    2013 & 14,038.17 &  5,476.24 &  2,261.80 & 1,600.27 \\
    2014 & 18,011.05 &  6,214.51 &  3,184.54 & 1,778.42 \\
    2015 & 23,919.52 &  7,000.64 &  3,571.18 & 1,981.28 \\
    2016 & 30,935.63 &  7,665.36 &  4,221.85 & 2,087.79 \\
    2017 & 40,059.40 &  8,889.58 &  5,367.63 & 2,513.28 \\
    2018 & 50,776.50 & 10,326.22 &  6,448.10 & 2,511.91 \\
    2019 & 63,116.82 & 11,461.56 &  8,856.39 & 2,598.47 \\
    2020 & 75,383.82 & 12,995.90 & 10,170.38 & 2,698.02 \\
    \hline
    \bottomrule
    \end{tabular}
    \caption{KLOC by source language for most recent release as of each year}
    \label{tab:loc_time_series}
\end{table}

\subsection{Other Examples}
Packages contain a wide variety of other information, including mandatory and optional fields in the required \textsc{DESCRIPTION} metadata file distributed with every package.  Many of these metadata fields actually drive behavior in the R interpreter.  As an example of deeper analysis that can be performed with our platform and data, we examine two additional types of metadata fields: those related to ``lazy'' data loading and those related to compilation of non-R source, e.g., Fortran, C, or C++. 

Table \ref{tab:lazy_compile_tag_summary} shows a breakdown of how many packages and releases tag Lazy Data, or as it also shows up in the metadata, Lazy Loading. Lazy Data is a reference to how a package handles the loading of data.  Normally, when R loads a package, it will read all objects into memory; when Lazy Loading is enabled, however, the interpreter will read only the data that the process currently needs instead of all package objects. This is a very useful flag for packages that may distribute large models or data files, and may be a good proxy to growth in recent data science applications within R. That over 60\% of packages have LazyData fields set to true does imply that a large proportion of packages vendor or contemplate vendoring data, but that there may be unexpected performance issues at execution time that result from such unexamined configuration.

\begin{table}[htbp]
    \centering
    \begin{tabular}{r|r|r|r}
    \toprule
    \hline
    {} &  LazyData &  NeedsCompilation \\
    \midrule
    \% of Releases &  58.27 &          20.52 \\
    \% of Packages &  63.20 &          17.95 \\
    \bottomrule
    \hline
    \end{tabular}
    \caption{Breakdown of Packages and Releases with LazyData and NeedsCompilation tags}
    \label{tab:lazy_compile_tag_summary}
\end{table}

To investigate this, we can look at the difference between the datasets included in these releases. Table \ref{tab:average_dataset_all} shows the breakdown for the average dataset included in an R release as compared to the average dataset included in a Lazy Data R release.  We can see that releases that are flagged as using lazy loading do have, on average, larger included datasets. However, the difference in their averages is only about 5 MBs in size. Further, while both show a large right-skew - meaning that there are many small datasets, which have the average brought up by especially large datasets, which makes sense as R is a language that is often used to deal with big data - the lazy loading datasets actually exhibit a more exaggerated right-skew, as they have a higher mean and a lower median. 

\begin{table}[htbp]
    \centering
    \begin{tabular}{r|r|r}
    \toprule
    \textbf{Statistic} &         \textbf{All packages} & \textbf{Lazy Loading packages} \\
    \midrule
    Count &       57,684.00 &            34,935.00 \\
    Mean  &      264,669.80 &           269,792.16 \\
    Standard Deviation   &    1,644,041.99 &         1,801,426.24 \\
    Minimum   &           75.00 &                75.00 \\
    25th Percentile   &        1,390.00 &             1,308.00 \\
    50th Percentile   &        9,289.50 &             8,130.00 \\
    75th Percentile   &       96,034.50 &            91,350.00 \\
    Maximum   &  102,473,361.00 &       102,473,361.00 \\
    \bottomrule
    \end{tabular}
    \caption{Descriptive Statistics of the number of bytes for the average R dataset, conditioned}
    \label{tab:average_dataset_all}
\end{table}

The Needs Compilation tag, on the other hand, is very straightforward. It tells whether or not the release in question needs to be compiled before it can be used. Table \ref{tab:lazy_compile_tag_summary} shows that there are more than 20\% of releases and roughly 18\% of packages that have the Needs Compilation tag flagged as true. Since, on average, more releases need compilation than packages, this seems to indicate that some subset of packages have many releases that require compilation. Likely, there are developers who provide source code releases - perhaps even as a regular 'developmental' or 'trunk' version.

\section{Conclusion and Future Work}
\label{S:conclusion}
To our knowledge, the results presented above provide the most comprehensive, longitudinal record of the R package ecosystem available.  We aggregate packages across not just two decades of CRAN, but also include nearly a decade of GitHub and Bioconductor.  Further, our methods and data include not just package-level metadata, but also normalized source control metadata and file-level information.  Within this sample, we find that the R package ecosystem has grown by orders of magnitude, but that recent history suggests a potential change in this trend.  We also find an extreme concentration of responsibility on a very small number of maintainers and organizations.  While we exclude policy discussion and normative questions from the scope of this publication, further discussion and analysis by the community is likely warranted.

We intend to present future work on two tracks.  First, we have hundreds of millions of records derived from parsing R, Fortran, C, and C++ sources with \textsc{antlr}, and we intend to provide further metrics related to conventions, quality, and security in the R ecosystem.  Second, we will present in separate research a comparative analysis of R's licensing and license network relative to Python, focusing specifically on potential risks to the community and broader adoption in academia and industry.

\section{Acknowledgements}
\label{S:acknowledgements}
This work was supported by Licensio, LLC, of which both authors are members.  The authors would like to acknowledge the late Professor Rick Riolo, whose spirit of inquiry into complex systems everywhere lives on in his many students around the world.

\bibliographystyle{model1-num-names}
\bibliography{empirical_r.bib}

\end{document}